\documentclass[10pt]{iopart}
\usepackage{graphicx}

\usepackage[dvipsnames]{xcolor}

\begin{document}

\title{Revisiting Atomic Collisions Physics with highly charged ions, A tribute to Michel Barat}

\author{Philippe Roncin}

\address{Institut des Sciences Mol\'{e}culaires d'Orsay (ISMO), CNRS, Univ. Paris-Sud, Universit\'{e} Paris-Saclay, b\^{a}t. 520, Orsay, France}
\ead{philippe.roncin@u-psud.fr}
\vspace{10pt}
\begin{indented}
\item[]April 2020
\end{indented}

\begin{abstract}
Michel Barat passed away in November 2018 at the age of 80 after a rich career in atomic and molecular collisions.  
He had participated actively in formalizing to the electron promotion model, contributed to low energy reactive collisions at the frontier of chemistry.
He investigated electron capture mechanisms by highly charged ions, switched to collision induced cluster dissociation and finally to UV laser excitation induced fragmentation mechanisms of biological molecules.
During this highly active time he created a lab, organized ICPEAC and participated actively in the administration of research.
This paper covers the ten years where he mentored my scientific activity in the blossoming field of electron capture by highly charge ions (HCI).
In spite of an impressive number of open channels, Michel found a way to capture the important parameters and to simplify the description of several electron capture processes;
orientation propensity, electron promotion, true double electron capture, Transfer ionisation, Transfer excitation, formation of Rydberg states, and electron capture by metastable states. Each time Michel established fruitful collaborations with other groups.
\end{abstract}

%
%
%
%
\ioptwocol

\section{Introduction}
In the 1960's he contributed to the birth of atomic collisions, he addressed an impressive number of different situations spanning from the construction of the electron promotion model in ion-atom collisions to low energy reactive collisions, collisions of atoms with ionic clusters, and fragmentation mechanisms of biological molecules induced by UV laser excitation.

Michel Barat always considered that if a problem is well-understood, theoretically or experimentally, then a simple minimalist model with a semi-quantitative predictive power can be derived. 
In other words an extensive calculation or a match between results and qualitative prediction was not an explanation. To reach this goal, he was very good at making collaborations and he created the laboratoire des collisions atomiques et mol\'{e}culaires (LCAM) as a place where both experimentalists and theoreticians could interact as often as possible. 
He was always completely open to collaborations around the world. 
I will try to describe this aspect by few examples, with apologies to the numerous other contributions that were made worldwide and that will not be quoted as they deserve, because this paper is more centered around the contributions of Michel Barat. 

The paper is organised as follow, the first section tries to recall briefly the context. 
The second one addresses specifically the technological challenges that we faced to investigate collisions with highly charged ions that can be skipped by readers only interested in scientific results. 
The following sections do not follow a chronological order but try to emphasize a specific problem to which Michel brought a significant contribution starting from single electron capture and followed by multiple electron capture.
The last section describes succinctly a few examples of experimental realizations that can be traced to the legacy of Michel.
\begin{figure}[ht]
\centering
\includegraphics{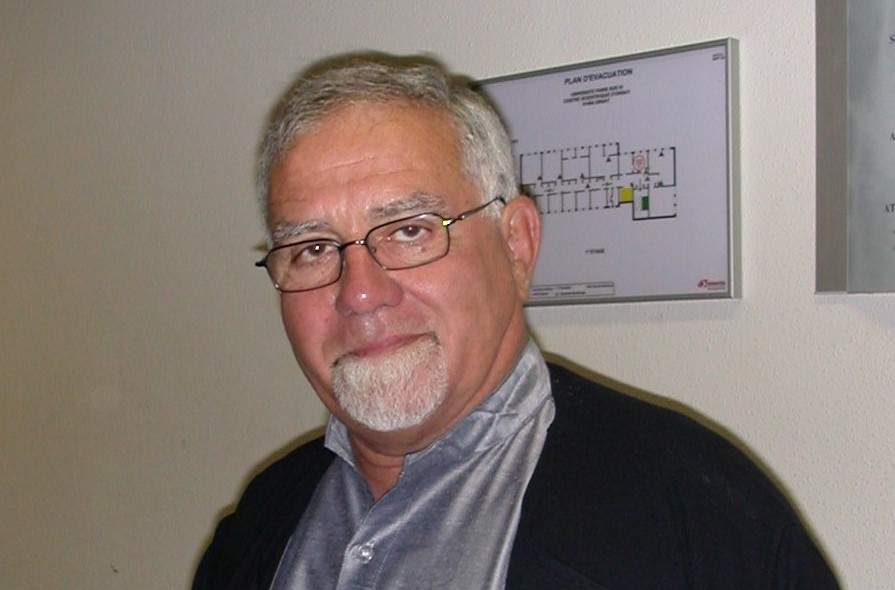}
\caption{Michel Barat played a key role to promote the physics of atomic and molecular collisions.}
\end{figure}

\section{Collisions with highly charged ions, a brief context}
\begin{figure}[ht]
\centering
\includegraphics[width=0.5\textwidth]{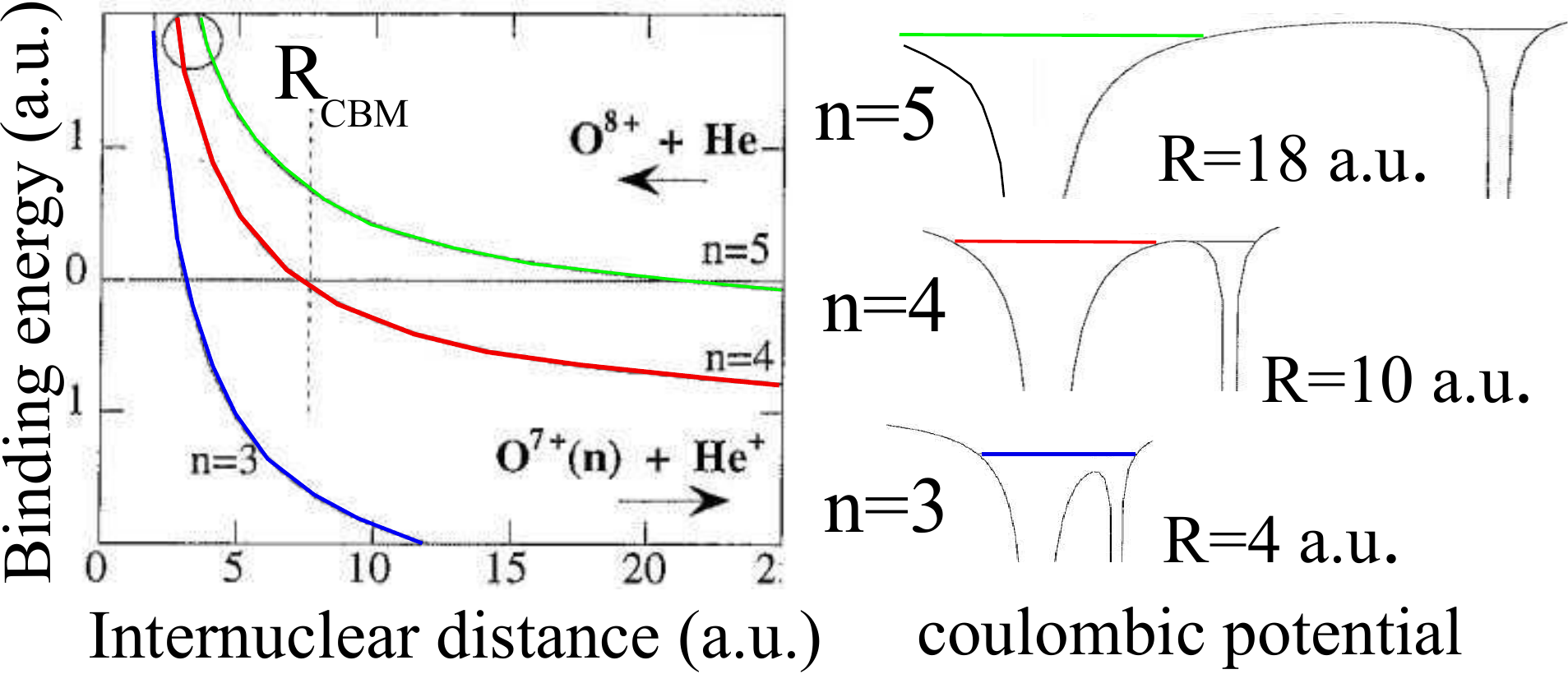}
\caption{The coulombic barrier model (CBM) considers the evolution of the coulombic barrier between target and projectile and points out the distance $R_{CBM}$ where the target electron can classically proceed toward the projectile. For the $O^{8+} + He$ system, the model indicates that capture should take place into the $O^{7+}(n=4)$ shell, the $O^{7+}(n=5)$ is too far away and separated by an very large barrier while the $O^{7+}(n=3)$ corresponds to a quasi molecular situation.
\label{fig:MCBM}}
\end{figure}

One of the great world-wide technological and scientific challenges was, and still is, the construction of civil nuclear fusion devices.
It was soon realized that impurities were present in the plasma as highly charged ions (HCI) and that many X-rays were emitted following electron capture by these HCI.
This was identified as one of the factors limiting the temperature increase of the Tokamak plasma above a million degrees.
\textit{In situ} optical diagnoses of the plasma was an important activity involving academic groups, (e.g. F.J. De Heer) whereas beam foil spectroscopy where MeV ions undergo numerous electron impacts when passing though thin carbon foils was used to identify spectral lines.
In both cases, quite a number of lines were present simultaneously and almost no experimental measurements of well defined collision processes were available to help modeling. 
Fortunately, the first ion sources able to produce highly charged ions were becoming available.
The most impressive, technologically, were the electron beam ion source (EBIS) invented and developed by E.D. Donets from Dubna in Ukraine.
Michel had started a collaboration with H. Laurent from the Orsay Institute of Nuclear Physics where J. Arianer was developing SILFEC, one of the first EBIS in the western world.
In Grenoble, R. Geller from the commissariat à l'\'{e}nergie atomique (C.E.A.), was investigating hot plasmas in a hexapolar magnetic structure heated by microwaves and demonstrated the presence of a broad range of highly charged ions.
This was the beginning of a new generation of ion source known as electron cyclotron resonance ion source (ECR).
S. Bliman, from Grenoble, and Michel managed to convince the C.E.A. to build a dedicated beam line and to open access to the academic community.
Pioneering work had already started with ideas around the coulombic barrier model (see in fig.\ref{fig:MCBM}) originally proposed by N. Bohr and J. Lindhard (1954). 
Adapted to highly charged ions this model outlined the key role played by the very large electric field attached to the projectile and pointed to the region of interest for electron capture .
The model clearly indicated that the situation was probably very different from that of singly charged ions where Michel had contributed to the development of the electron promotion model \cite{Barat_1972}.
Many groups around the world started to imagine experiments to probe these new exotic highly charged ions. 
The first international meeting of this growing community around highly charged ions took place in 1982 in Stockholm.

\section{Technological challenges}
\begin{figure}[ht]
\centering
\includegraphics[width=0.5\textwidth]{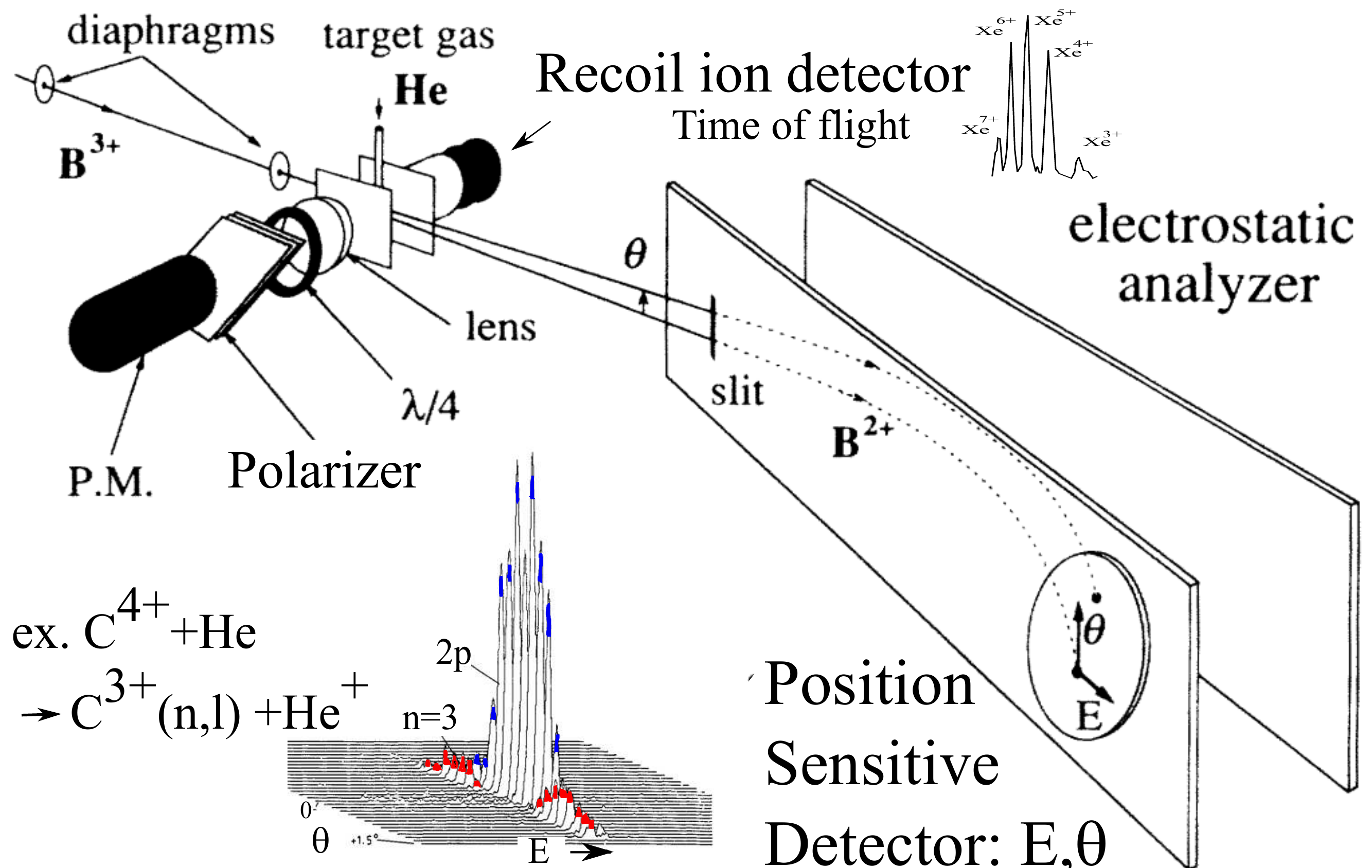}
\caption{Scheme of the experimental setup. The electrostatic analyser selects the projectile final charge state and the P.S.D. records its energy gain and scattering angle in coincidence with recoil ions.
\label{fig:setup}}
\end{figure}

Michel has always been interested in technological developments that could help his experiments and was not afraid of a challenge. 
I started my PHD in 1983, by assembling the electrostatic spectrometer (Fig.\ref{fig:setup}) for translational spectroscopy that Michel had designed. 
It was a simple but large 45$^\circ$ parallel plate electrostatic analyser designed to host a 2D position sensitive detector. 
The only snag was that these detectors were not available at that time! Micro-channel plates developed by the military industry for night vision application were indeed becoming available but quite expensive. 
Copying a design from D. de Bruyn of the F.O.M. Institute in Amsterdam \cite{Knibbeler_1987}, I started by constructing a 1D version based on the capacitive charge division technique :
A set of 50 discrete copper anodes printed on fiberglass substrate and interconnected by capacitors. 
If the electron cloud of charge Q$\approx 10^7 e^-$ produced by the two MCP is narrow and falls entirely on the $i^{th}$ anode, then the charge Q will split according to the number of capacitors to be passed before reaching the leftmost or rightmost anode allowing identification of the anode i; $i=\frac{50 \times Q_{left}}{(Q_{left}+Q_{right})}$.
On the contrary, if the electron cloud is broad enough, the system rapidly points to the charge barycenter and provides a continuous localisation $x=\frac{Q_{left}}{(Q_{left}+Q_{right})}$. 
H. Laurent provided state of the art analog electronic modules; preamplifiers and linear pulse shaping amplifiers as well as specific modules to perform the analog addition and analog division required to obtain the position $x$ of the impact. 
With help of L. Lilieby during a short visit from Stockholm, we managed to measure our first energy gain spectra with ions extracted from SILFEC. 
Michel immediately pushed me to develop a 2D version following a suggestion by D. De Bruyn (F.O.M. Amsterdam).
It was made of two 1D arrays as above but with a triangular shape inter-penetrated anodes similar to a game of backgammon.
We then faced the difficulty of performing so many analog additions and divisions; each step being sensitive to noise and to thermal drift. 
We also tried to perform division inside our brand new BFM186, the first 8086 personal computer at the lab, but even with integer divisions programmed in assembly language the acquisition rate was below a kHz. 
J.C. Brenot, from LCAM and expert in micro electronics convinced me that making a machine that would control all ADC's and perform the division would be simple. 
His argument being that a 12 bits division is nothing else that twelve subtractions, "if you hard wire the carry bit to a shift register and reload the result, it can be done in 12 clock cycles".
I was naive, supported by Michel I started assembling (wrapping) a dedicated micro computer around a 12 MHz clock, an AMD2910 sequencer driving 5 chips of static RAM memory chips configured in 128 lines of 40 bits. Each line had 10 bits dedicated to the sequencer 3 bits for instructions plus 7 bits to drive the above 128 lines of memory address, few bits for an arithmetic and logic unit and few bits to manage the combined operation of the four ADC's. 
I had a hard time getting it to work properly but I learned to check the logic, the propagation delays, to track glitches and read documentation over and over... I learned micro electronic from the oscilloscope.

It is very difficult to imagine now, the acquisition program inside our PC was written in assembly language without connection to the newly created MS-DOS operating system.
Our "micro-divider" was connected directly to the interrupt priority controller forcing the processor to execute the acquisition program at each event. 
No need to say that every tiny programming mistake usually resulted in the PC crashing..
But it worked and the acquisition was able to sustain 10 kHz rate including the generation of 2D plots written directly in the graphic memory mapping the green screen.

At this moment, other groups already succeeded to perform energy loss measurements : P. Hvelplund in Denmark, H. Tawara in Japan, B. Huber in Germany, L. Coke in the USA  and R. Mc Cullough in U.K. 
We did not have the best resolution but the simple fact that we did not have to scan the energy or the scattering angle made our setup extremely efficient, a decisive advantage for the limited time available at external facilities.
The large efficiency is also particularly useful for experiments where the recoil ion or the emitted electrons and photons have to be detected in coincidence because the overall efficiency is the product of both the individual detection and collection efficiency.
Michel had chosen this technique because it is sensitive to the primary processes taking place at short inter-nuclear distance and is much less affected by the very complex decay processes of the highly excited states formed after electron capture.


\section{In what sense is the captured electron rotating ?}
In Denmark, N.O. Andersen had started to derive propensity rules for alignment and orientation in atomic collisions \cite{Andersen_1987} and D. Dowek, a former PHD student of Michel had also constructed a 2D detector to study these effects with laser excited targets \cite{Dowek_1990}. 
After discussions, Niels and Michel selected the $B^{3+}(1s^2)$ + $He$ system to investigate the orientation rule ; would the $B^{2+}(1s^22p)$ electron rotate clockwise or anticlockwise?
The scattering profile would give access to the impact parameter, the energy gain would identify the $B^{2+}(1s^22p)$ capture level and the circular polarisation of the 207 nm $B^{2+}(2p \rightarrow 2s)$ decay photon would indicate the rotation sense of the captured electron.
\begin{figure}[ht]
\centering
\includegraphics[width=0.4\textwidth]{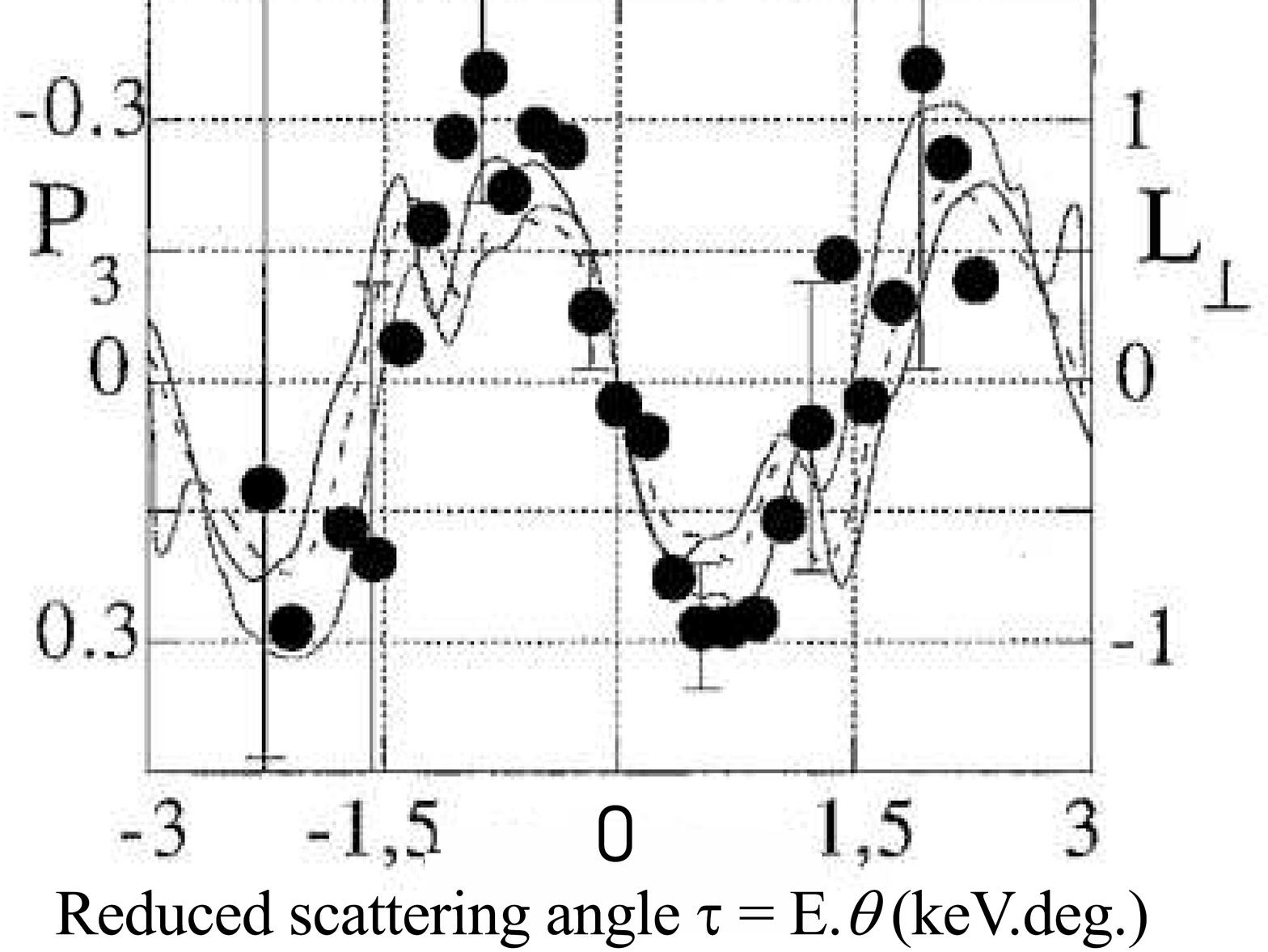}
\caption{Taking into account the polariser efficiency, the polarization $P_3=\frac{RHC-LHC}{RHC+LHC}$ measured at 1.5 keV \cite{Roncin_1990} is transformed into the angular momentum $L_\perp$ reaching one unit. Lines are for an extensive theory \cite{Roncin_1994} and a simple collision model \cite{Adjouri_1994,Ostrovsky_1991}.
\label{fig:B3_pol}}
\end{figure}
We designed and constructed a large aperture optical system mounted in our setup (fig.\ref{fig:setup}) to allow a triple coincidence measurement between the photon, the recoiled $He^+$ and the scattered $B^{2+}$ ions.
Together with C. Adjouri (PHD), we were lucky to succeed on the first run in Grenoble \cite{Roncin_1990}. 
The measured orientation displayed in fig.\ref{fig:B3_pol} shows that the electron is turning in the "intuitive" sense.  
The experiment indicated that it is easier to jump on a rotating merry go round if you run along side such that the relative velocity is reduced. This was was a bit puzzling because this should not hold in the molecular regime where the electron motion is much faster (here up to 20 times) than that of the nuclei. 
Our observations were first reproduced by J. P. Hansen, A. Dubois, and S. E. Nielsen \cite{Hansen_1992} in a state of the art close coupling calculation.
Then, V.N. Ostrovsky from Saint Petersbourg that Michel had invited during the difficult period of the collapse of soviet union, derived an analytical model based on a Landau-Zener description adapted to circular states \cite{Ostrovsky_1991}.
The explanation he had found was simple enough to be part of textbooks. 
In the genuine two-states Landau-Zener model electronic transitions take place at localised crossing between the energy curves with a probability $p$.
In a collision, the crossing has to be passed twice, on the way in (while approaching the target) and on the way out giving two different possibilities ; capture on the way in and not on the way out with a probability $p\times (1-p)$ or the reverse with an equal probabilities $(1-p)\times p$.
A curve crossing indicates that both states have a different energy dependence so that the semi classical phase developed along these paths is different and evolves with the impact parameter producing the well-known Stuekelberg oscillations in the differential cross section of a simple system e.g. He$^+$ +He$\rightarrow$ He +He$^+$.

Here the situation is different, there are three $B^{2+}(1s^22p)$ states : $2p_x, 2p_y$ and $2p_z$. 
By symmetry only the linear combination $2p_\sigma$ aligned on the inter nuclear (molecular) axis is available for electron capture.
If it is strongly coupled to the molecular axis, then after rotation, the same $2p_\sigma$ state will be available for capture (or not) on the way out and nothing is changed, Stuekelberg oscillations will be observed.

\begin{figure}[ht]
\centering
\includegraphics[width=0.45\textwidth]{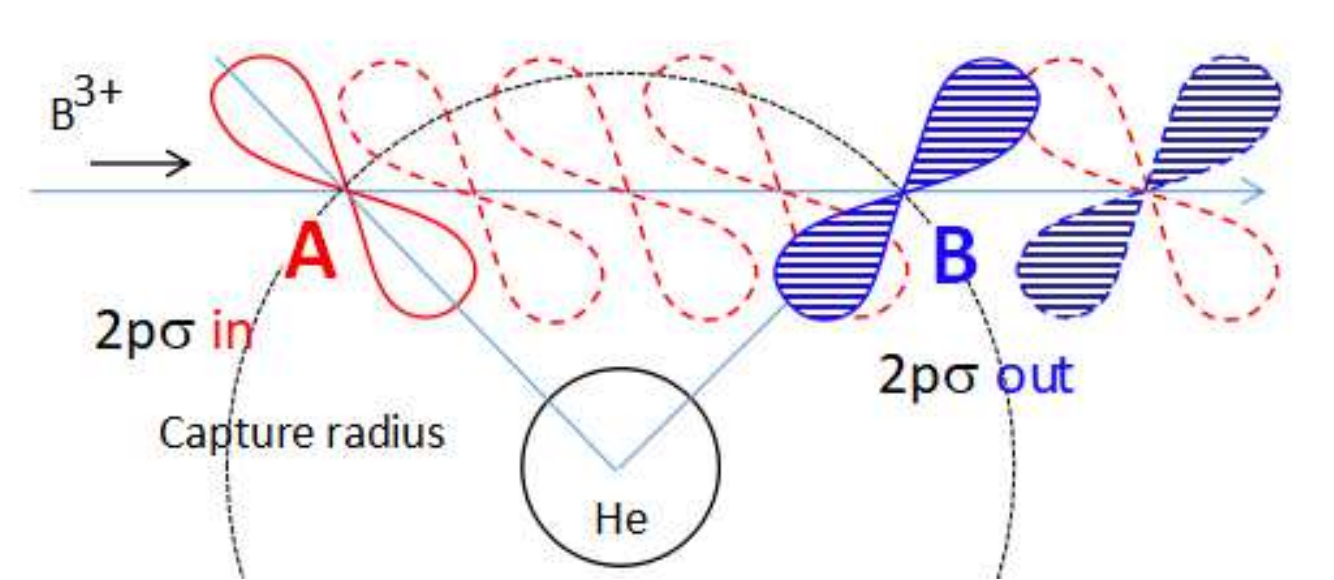}
\caption{schematic view of the $B^{3+}(1s^2)$ + $He$ collision, the \textcolor{red}{$2p_\sigma$} state populated at the first crossing \textcolor{red}{\textbf{A}} tends to remain fixed in space so that it hardly interacts at the second crossing \textcolor{blue}{\textbf{B}}. In this example, a phase difference of $\pi/2$ gives a circular state \cite{Ostrovsky_1991}.
\label{fig:B3_model}}
\end{figure}
In contrast, in a diabatic situation where the $2p_\sigma$ capture state is hardly forced to follow the rotation of the molecular axis, then  the $2p_\sigma$ state available on the way in, tends to stay fixed in space and, on the way out, it is likely to look like a $2p_\pi$ state in the molecular frame,  whereas symmetry still imposes capture into the $2p_\sigma$. 
The "Stueckelberg" phase difference now applies to different states as depicted in fig.\ref{fig:B3_model} and produces an oscillation between linear and circularly polarized states.

With this model in mind and after illuminating discussions with V. Sidis we could calculate electron capture coupling elements and build a three state model reproducing the observed polarization \cite{Adjouri_1994}.
The intuitive sense of rotation was indeed purely accidental.
In a way, this corresponds to a situation where rotational coupling was quite strong whereas the quasi-molecular description often assumes an absence of rotational coupling.
Note that, in the mean time, alignment \cite{Dowek_1990} and orientation measurements with laser excited alkali targets had developed to establish propensity rules from the reversed process \cite{Salgado_1997}.

\section{Why do we observe higher $n$ value?}
The selective electron capture into a given $n$ shell was well explained by the coulombic barrier model but both electron and photon spectroscopy provided evidence for electron capture into a higher lying $n$ shell. 
These were particularly important for optical diagnostic of plasma because decay in the V.UV. range is more likely for large $n$ values.
Within the CBM, the crossing associated with direct electron capture corresponds to a tunneling situation through a very large barrier separating the projectile and target core (top curve in fig.\ref{fig:MCBM}).
The cross section should be vanishingly small and the differential cross section peaked at very small scattering angle.
With S. Ohtani, on leave from Nagoya, we observed the opposite : large cross sections peaking at large scattering angles \cite{Roncin_1990np1}.
Our data pointed to an indirect sequential process where an electron is first captured into the shell $n$ predicted by the CBM on the way in, and then, at smaller inter-nuclear distance, transferred to the $n+1$ shell. For $O^{8+} + He$, this indicated a first capture into $O^{7+}(n=4)$ followed by a coupling to $O^{7+}(n=5)$ at shorter distance as depicted by the circle around $R\approx 3-4$ a.u. in fig.\ref{fig:MCBM}.
It did not take long for Michel to recognize a situation that had not yet been observed, only predicted by the Fano-Barat-Lichten electron promotion model \cite{Fano_1965,Barat_1972}. 
As illustrated in fig.\ref{fig:MCBM}, at short distance, the separated atoms is not adequate because the excited electron evolves overs both the target and projectile core.
The coulombic barrier model provides an adequate indication of the favorable capture distance but it only considers the inter-nuclear coulombic repulsion.
\begin{figure}[ht]
\centering
\includegraphics[width=0.5\textwidth]{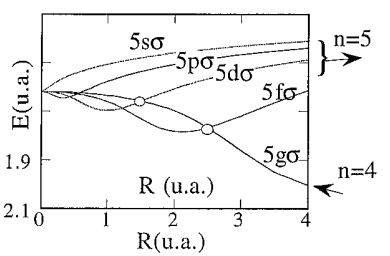}
\caption{schematic view of the $O^{8+} + H$ molecular orbitals at short inter-nuclear separation. The different shells of the separated atoms have to converge towards the energy levels of the united atom $(R=0)$. One $\sigma$ state the $5g\sigma$ is promoted and crosses the $5d\sigma$ and $5f\sigma$ states correlated to the $n=5$ shell of the separated atoms.
\label{fig:OEDM}}
\end{figure}
Let us forget this contribution and focus on the weaker electronic terms for a one electron system as displayed in fig.\ref{fig:OEDM}. The general solution can be expressed in terms of elliptic states but
one can distinguish two asymptotic quasi atomic regions, the separated atoms when the inter nuclear-distance is larger than the size of the electronic wave-function and the united atom when it is much smaller.
At zero inter-nuclear separation, an excited electron would see a single nucleus of charge 9, and as the nuclei start to separated; the non spherical nuclear charge distribution gives rise to the quadrupolar splitting (left of fig.\ref{fig:OEDM}) slowly converging to stark states of the separated atoms when the recoiled $H^+$ ion polarizes the $O^{7+} (n=5)$ ion at large internuclear distance.
Since the number of states are different one state of the separated atoms has to be promoted to the $n+1$ shell of the united atom.

\section{Are the electrons captured simultaneously or one after the other?}
Our doubly differential measurements were well-adapted to address this naive question.
First, for a given target atom, we measured rather similar energy gain and scattering profiles for projectiles with identical charge state $C^{6+}, N^{6+},O^{6+},F^{6+}$, or $N^{7+},O^{7+},Ne^{7+}$ or $O^{8+},Ne^{8+}$ \cite{Roncin_1986b,Barat_1987} indicating that the ion core structure does not play a key role in the primary capture mechanism whereas it has a drastic influence on the fate of the multiply excited ion after the collision; its final charge state, the emitted electron spectra as well as Xray or V.UV spectra.
Focusing on double electron capture, we observed both one step and two step double electron capture so we reformulated the question differently.
What is the driving mechanism in simultaneous double electron capture? 
More precisely, does the simultaneous capture of two electrons at a curve crossing imply that the di-electronic $1/r_{12}$ interaction is responsible?
When the two electrons are captured one after the other, each crossing can be associated with a one electron exchange term but this is not the case in simultaneous double electron capture where a second order term has to be considered. 
The question was therefore are second-order one-electron-exchange terms more important than first order di-electronic terms?
With the help of V. Sidis, we decided to investigated this theoretical aspect on the $C^{4+}+ He$ system considered as the ideal candidate because not only double electron capture takes place in one step but it is more important than single electron capture as measured \textit{e.g.} by Okuno \textit{et al} \cite{Okuno_1983} and confirmed theoretically by Kimura and Olson \cite{Kimura_1984}.
By reproducing the measured angular dependence of all observed channels semi-quantitatively, we could show \cite{Barat_1990} that double electron capture can be explained by a second order effect without the need of di-electronic interaction as an electron capture mechanism.
This early work was confirmed by later more extensive and more detailed studies \cite{Hansen_1992b,Errea_1995,Gao_2017}.  
The lesson is not that the di-electronic interaction is unimportant as will be discussed later in the study of the evolution of the multiply excited states, but that it is not the dominant driving force for electron capture \cite{Barat_1993}.
The importance of second order effects can be understood from the CBM model in fig.\ref{fig:MCBM}, if the inter-nuclear distance $R_{CBM}$ favorable for single electron capture is passed without capturing an electron, the level of the target electron is now above the saddle point and free to explore the projectile. 
Describing these states in an atomic basis requires handling a large basis set beyond the implicit first order perturbation theory.
It is not completely surprising that a description in terms of well separated projectile and target wave functions is hardly relevant below $R_{CBM}$ \cite{Roncin_1987}. 

The same quasi-atomic description with second order perturbation theory, was found to describe very well other two-electron process such as core changing electron capture by HCI, also called Transfer-Excitation, where two electrons are involved:

$Ne^{7+}(1s^2 2s) + He \;\rightarrow Ne^{6+}(1s^2 2p 3l) + He^+$ \cite{Roncin_1986c}

$F^{6+}(1s^2 2s) + He \;\rightarrow F^{5+}(1s^2 2p 3l) + He^+$ \cite{Gaboriaud_1994}

$C^{4+*}(1s 2s\; ^3S) + He \;\rightarrow C^{3+}(1s 2p^2) + He^+$ \cite{Guillemot_1990b}

Theses results were extended to multiple electron capture by L. Guillemot \textit{et al} \cite{Guillemot_1990,Barat_1992} during his PHD and can be also formulated in terms of characteristic times; the typical time $\tau_{e-e}$ for di-electronic interaction can be as short as few hundred  atomic time units (1 a.u. $\sim 2.42~10^{-17} s$), fast enough to have have some influence during the collision \cite{Stolterfoht_1986} but much slower that the the time $\tau_{capt}\sim~1$ a.u. needed for one electron to pass the barrier.

\section{How do highly excited electrons manage to avoid autoionisation?}
One of the puzzling results was that after multiple electron capture many of the multiply excited states managed to decay by photon emission, avoiding the much faster process of autoionisation.
First, only measurements charge states measurements were available, by coincidence with doubly charged recoil target ions, one selects projectile ions that have captured two electrons, and the final charge states indicates if the projectile managed to keep them both (true double capture) or if one electron was released by autoionisation (auto-ionizing double electron capture). 
The ratio of true double capture was called stabilization ratio and was first expected to reflect an average branching ratio for radiative decay of the doubly excited states populated by the collision.
With our spectrometer we could bring energy resolution and identify the principal quantum numbers involved in the primary collisional processes.
With the group of A. Ch\'etioui we performed triple coincidences with emitted X-rays to investigate the possibility that specific doubly excited states could be formed having a reduced Auger rate.
Our results were compatible with branching ratios calculated by F. Martin but did not show sufficient contrast to derive any conclusion \cite{Chetioui_1990}.
We also tried a large core system such $Ar^{7+} (nl, n'l')$\cite{Guillemot_1990b} with the hope that so many different states would be populated that maybe some trend could appear by comparison with transition arrays calculated by E. Luc and J. Bauche \cite{Luc_Koenig_1990}.
Here again no clear result appeared, but we now understand that we made a mistake. After discussion with the theoreticians, the final step of massive configuration interaction between the thousands of doubly excited states was abandoned.
We also had to face the profound contradiction that electron spectroscopy \textit{e.g.} by A. Bordenave-Montesquieu \textit{et al} \cite{Bordenave_Montesquieu_1984}, M. Mack \cite{MACK_1987} or N.Stolterfoht \textit{et al} \cite{Stolterfoht_1986} was not reporting the same selectivity of the  electron capture processes. 
They were observing only a comparatively weak population of the quasi-symmetrical doubly excited states $(n, n'\simeq n)$ that we were observing as primary populated states in agreement with coulombic barrier model.
We have to mention that another successful interpretation of the coulombic barrier model by A. Niehaus \cite{Niehaus_1986} stated that electrons are "molecularized" on the way in, at a given binding energy and redistributed among all possible states. 
The reaction window derived from this model was found to match the broad range of doubly excited states observed in electron spectroscopy \cite{Mack_Niehaus}.

The situation became clear when we investigated the systems $Ne^{10+} + He$, $Ne^{8+}$ or $O^{8+} + Ne$, $N^{7+}$ or $O^{7+} + Ar$ where an important population of $(4l,4l')$ and $(4l,5l')$ doubly excited series was observed.
The  $(4l,5l')$ series were all found to decay dominantly by auto-ionisation while the $(4l,4l')$ series would decay mainly radiatively and therefore "stabilise" the two captured electrons on the projectile \cite{Roncin_1991}. 
Obviously the naive interpretation that two electrons in the same shell interact more strongly (auto-ionize faster) than if they are in different shell was not the correct answer.

\begin{figure}[ht]
\centering
\includegraphics[width=0.35\textwidth]{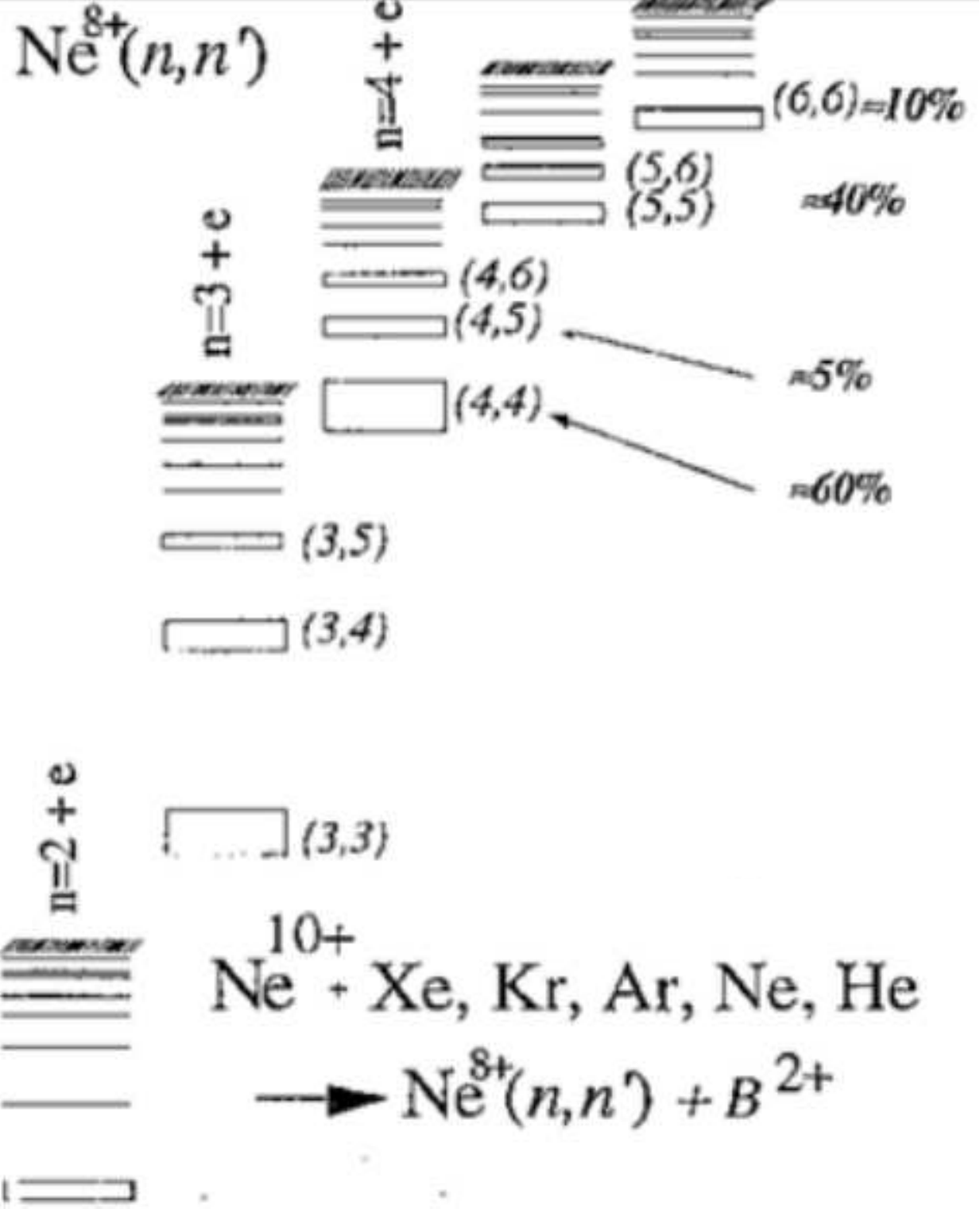}
\caption{Low keV energies collisions of $Ne^{10+}$ ions with $Xe,\; Kr,\; Ar,\; Ne$ and $He$ target indicate an initial population of several quasi-symmetrical double excited series ranging from $(4,4)$ to $(6,6)$. The percentage indicated next to the series reports a typical stabilization ratio. Doubly excited states lying close and below the top of Rydberg series hardly auto-ionize \cite{Roncin_1991}.
\label{fig:straddling}}
\end{figure}
By coincidence with doubly charged recoil target ions, we select projectile ions that have captured two electrons and the question is how many will manage to keep them both (true double capture) and how many will release one by autoionisation (auto-ionizing double electron capture). The ratio of true double capture was called stabilization ratio and is reported in fig.\ref{fig:straddling}
Fig. \ref{fig:straddling} reports stabilization ratios observed for doubly excited series identified as initially populated for various systems.
The largest stabilisation ratios are observed when the energy of the doubly excited series $(n,n'\simeq n)$ are lying below the top of an adjacent Rydberg series.
Hence the following step wise decay, the $Ne^{8+} (4l,4l')$ would dilute into the energy straddling $Ne^{8+} (3l,n'l')$ series which is then believed to be sufficiently asymmetric to avoid auto-ionisation.
The interaction diluting an initially populated $Ne^{8+} (4l,4l')$ state is nothing else than the di-electronic $1/r_{12}$ interaction responsible for dilution in the auto-ionisation continuum that would be available if the state would have been located above the $Ne^{8+} (3l,n\sim \infty)$ limit.
In line with the quantum defect theory, we proposed with H. Bachau to calculate the decay width as if the autoionisation would be allowed \cite{Bachau_1992}.
The matching in energy does not have to be precise because, due to incomplete shielding of the nuclear charge by the Rydberg electron, the relative energy position evolves with inter-nuclear distance in exactly the same way as it does for states auto-ionising at finite distance from the projectile so that the line profile is affected by the so called post-collisional interaction.
The importance of a Rydberg series close to the initial capture levels was confirmed by other groups using ion \cite{Ali_1993}, electron\cite{Bordenave_1994} and photon spectroscopy \cite{Martin_1994}.
However, when trying to make this model quantitative by taking calculated partial transfer width to well-defined spectroscopic terms, we faced the difficulty that the populated $Ne^{8+} (3l,n'l')$ series were not auto-ionizing fast enough to explain the observations \cite{Vanderhart_1994}.
This is because momentum conservation indicates a low $l\ll n$ value for the Rydberg electron, in other words, the $Ne^{8+} (3l,n'l')$ Rydberg electron leaving the $Ne^{8+} (4l,4l')$ core will return close to the $n=3$ core after a Rydberg period and have a new chance to auto-ionize.
Somehow, the orbit of the Rydberg electron has to be moved away from the core.

\begin{figure}[ht]
\centering
\includegraphics[width=0.4\textwidth]{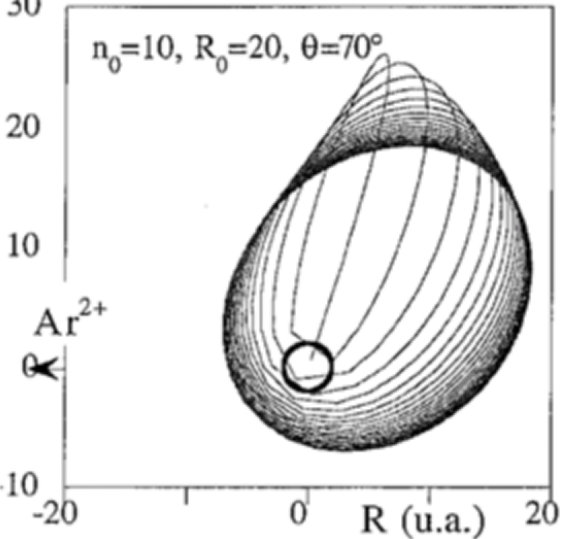}
\caption{Classical trajectory corresponding to a $Ne^{8+}(3,n=10)$ Rydberg electron emitted in the field an $Ar^{2+}$ receding target ion initially at a distance R=20 a.u.
\label{fig:Rydberg_PCI}}
\end{figure}

Once again, we were lucky to have a skillful visiting scientist on leave from Saint Petersburg;
A.K. Kazansky not only understood the problem before us, but proposed an analytic estimate of the stark mixing in the receding target field\cite{Kazansky_1992}.
In most situations, the angular momentum of the Rydberg electron was found to increase \cite{Kazansky_1994}, as expected from phase space considerations, explaining the observed quenching of auto-ionisation \cite{Roncin_1993,Roncin_1997}.
This also explained in more detail the stepwise $\delta n=1$ Rydberg transitions from circular Rydberg states (\textit{i.e.} the Yrast cascade) observed by S. Martin \textit{et al} \cite{Martin_1990} as well as the equivalent lines emitted by the receding excited target ion.
Some of the Rydberg electrons ejected by the projectile were simply re-captured by the target as easily imagined from fig.\ref{fig:Rydberg_PCI} if the initial direction points to the target.
We combined our observations with those of the Lyon group to better describe the population mechanism of double Rydberg series deriving from the dilution of quasi symmetrical triply excited series populated in collisions of highly charged $C^{6+}$ and $Ne^{10+}$ ions \cite{Bernard_1997}.

Around 1994, we stopped our experiments in the gas phase with the satisfaction that we had managed to extract from the complexity of highly charged ions, a few examples of collisions that were even more simple than with singly charged ions. 
A with many students, I started my PHD with the impression that I would never be able to catch up with all the knowledge accumulated by generations of smart physicists. 
Thanks to Michel I had the feeling that, in our narrow field of HCI, I had revisited ion-atom collisions.
In less than ten years Michel \textit{et al} published more than 40 papers, always with great enthusiasm.

Then our scientific activity took different directions.
Other topics had developed such as testing of quantum electrodynamics \cite{Indelicato_1989} particularly in Livermore where they had transformed the EBIS source into an EBIT trap \cite{Levine_1989,Beiersdorfer_2003,Dragani_2003}, the hollow atom formation at surfaces \cite{Briand_1990,Limburg_1995,Arnau_1997,Winter_1999}, the guiding of highly charged ions through nano and micro capillaries \cite{Stolterfoht_2002,Ikeda_2006,Skog_2008,Cassimi_2009,Gruber_2012}. Many other groups were also equipped with their own ECR or EBIS ion source and by then, our setup was no longer unique. 

A new generation of setups with large time sensitive imaging detectors to correlate the impacts of several fragments from a collision \cite{Moshammer_1996,Lafosse_2000} was developing.
The most efficient "collision microscope" named COLTRIMS took advantage of cold target atoms from supersonic expansion to measure the energy balance and scattering angles, not from the projectile as we were doing but from the target recoil momentum and continued revisiting the basic collisional processes \cite{Moshammer_1996,Mergel_1995,Ullrich_1997,Flechard_1997,Jahnke_2004,Yan_2017}.

With J. Fayeton at LCAM, Michel started a new career in the emerging field of clusters by investigating atomic collisions with sodium clusters. 
Together with Y. Picard from LCAM and I. Ismail (PHD), they constructed a "zero dead time" position sensitive detector \cite{Ismail_2005}.
Ten years after Michel Barat started a last career on collision induced and laser induced dissociation of bio molecules. 
Together with C. Jouvet from Laboratoire de photo physique mol\'{e}culaire and J. Fayeton, Michel constructed "Arc en Ciel"  to investigate the fate of large bio molecules after the impact fs laser or the collision with atoms \cite{Jouvet_2005}.
Michel succeeded to have five fruitful careers, ion-atom collisions, reactive collisions, electron capture by HCI, collisons with cluster and bio-molecules, all related to collision physics with significant instrumental developments.

\section{My legacy from Michel}
\begin{figure}[ht]
\centering
\includegraphics[width=5cm]{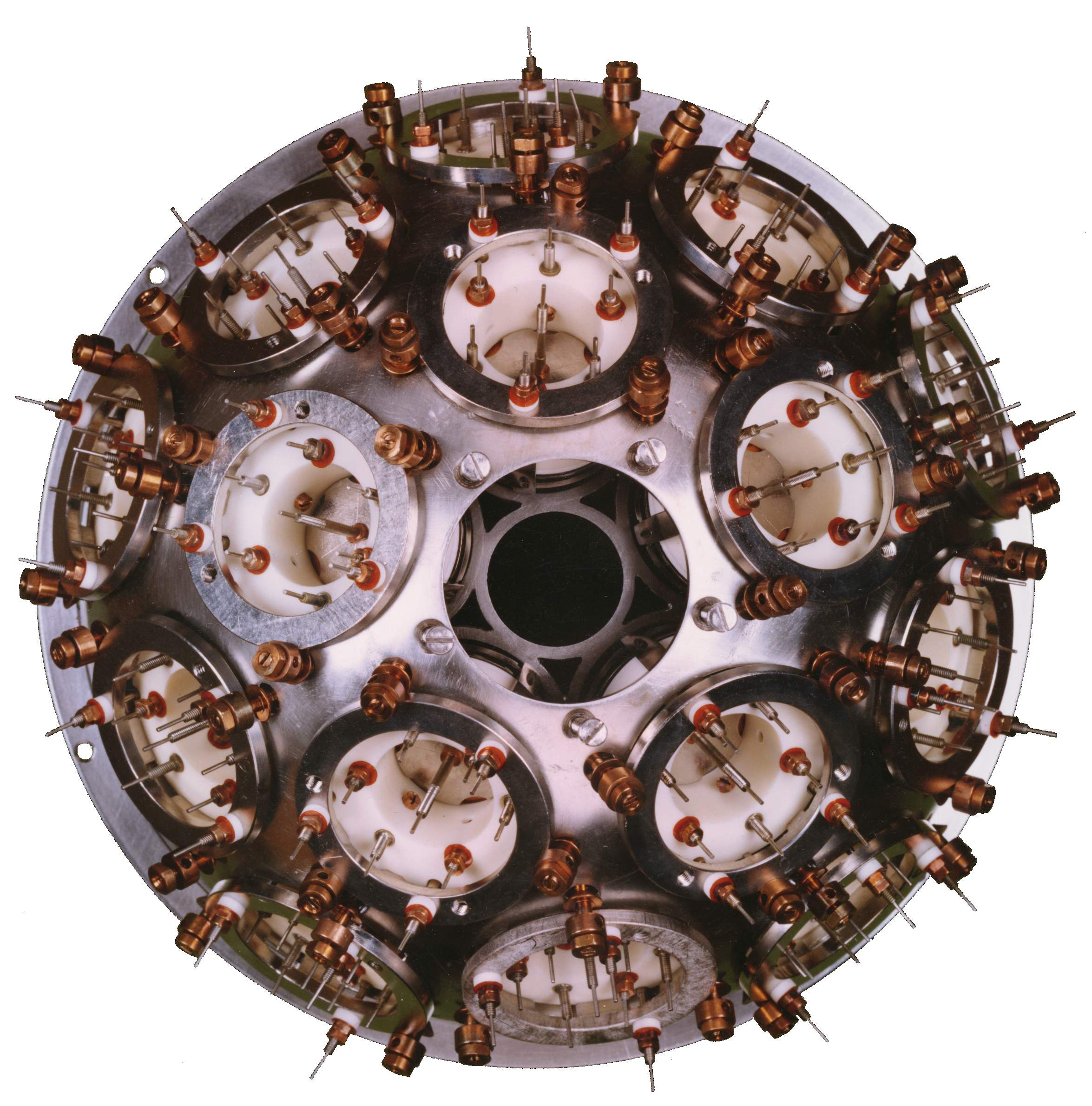}
\caption{Top view of the $2\pi$ multi-detector designed and constructed by V.N. Morosov. 15 independant MCP detectors are facing the center where the target surface is placed. Collisions could be investigated at grazing or normal incidence.
\label{fig:det_2pi}}
\end{figure}

Supported by Michel and J.P. Gauyacq, I switched to investigate atomic collisions at surfaces, in particular the impact of a single HCI had been found to convert their large potential energy into the emission of up to several hundred secondary electrons \cite{Kurz_1994} during the decay of the hollow atom \cite{Briand_1990,Limburg_1995,Arnau_1997,Winter_1999}.
I started designing a detector to collect the electrons emitted from the surface in coincidence with the scattered projectile, but once again, it was Michel who initiated discussions with Pr Leonas from I.P.M. Moscow. 
In close collaboration with V.N. Morosov and A. Kalinin from Moscow and Z. Szilagi (PHD), we constructed an ambitious $2\pi$ multi-detector \cite{Morosov_1996}.
The detector shown in fig.\ref{fig:det_2pi} was so efficient that, after calibration \cite{Villette_2000} it was easy to correlate quantitatively the presence or absence of a specific feature such as the detection of a fast Auger electron with the number of slow electrons \cite{Morosov_1997}.

We decided to apply this detector to investigate the basic electronic processes taking place during collisions at surfaces.
With PHD students, J. Villette and J.P. Atanas, we started to address the paradox that, under ion impact, insulator surfaces tend to emit more electrons than metal surfaces in spite of a work-function up to twice as high.
Though each projectile undergoes several successive collisions with the surface atoms, we succeeded to observe individual inelastic processes and, with help of coincidences, we could identify the collisional processes governing electron emission in collisions of keV protons on a $LiF(100)$ surface \cite{Roncin_1999}.

\begin{figure}[ht]
\centering
\includegraphics[width=7.5cm]{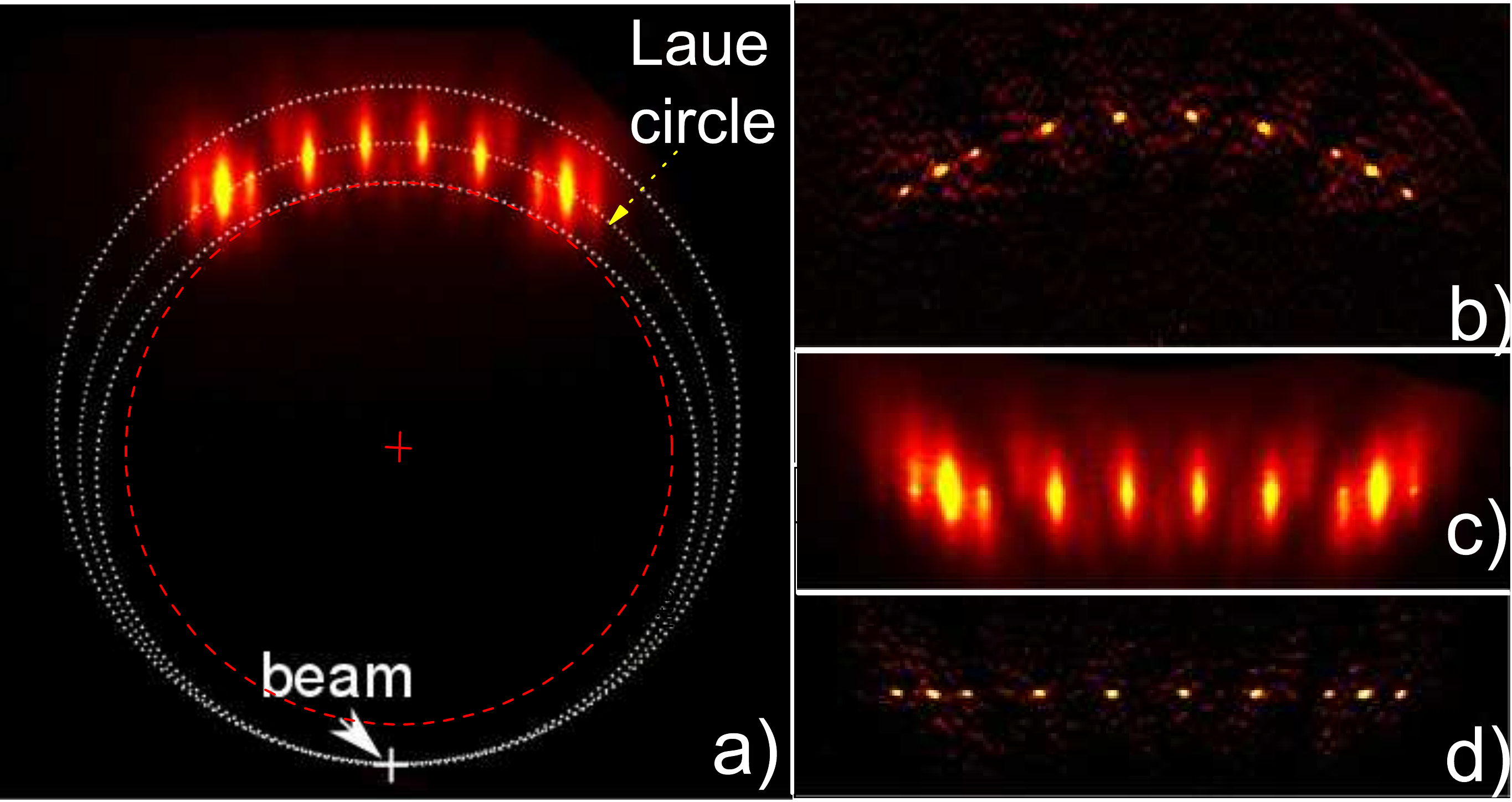}
\caption{a) scattering profile of 460 eV helium atoms impinging the surface of a LiF crystal at an incidence of 1.57 deg. Diffraction is observed and b), d) isolate the elastic contribution on the Laue circle where all surface atoms appear frozen at equilibrium position while c) shows the inelastic contribution where binary collisions have excited phonons \cite{Roncin_2017}.
\label{fig:diffract}}
\end{figure}

The incoming proton neutralizes before hitting the surface and then, at each alkali site, the $H^\circ$ atom has a chance to capture one electron by the ion-pair-like mechanism just discovered by C. Auth, A.G. Borisov and H. Winter \cite{Auth_1995}. 
We could show that when leaving the alkali site where capture took place, the $H^-$ ion has to cross the localised exciton level and that, most often, an $H^\circ$ atom would emerge because the captured electron was placed in this exciton level. 
The $H^-$ ions that could leave the alkali site with the captured electron were exposed to very efficient electron detachment when passing on top of the next alkali site, emitting a low energy electron and returning to the $H^\circ$ state from which new cycles of capture and loss could restart.
With J. Villette, A. Momeni (PHD), H. Khemliche  and A.G. Borisov both from LCAM, we could show that $Ne^+$ ions neutralize on the $LiF(100)$ surface via a dark-Auger process, without emitting electrons \cite{Khemliche_2001}. 
Just like the quenching of auto-ionisation by transfer to Rydberg series observed with Michel, the  $1/r_{12}$ di-electronic interaction was coupling the $Ne^*$ excited states to a different continuum, the LiF conduction band.
No electron was emitted because the excited $Ne^*$ state is coupled to a different continuum, the conduction band.
In the same spirit as the $C^{4+} + He$ system investigated with Michel \cite{Barat_1990}, we could identify a simultaneous double electron capture by $F^+$ ions forming a $F^-$ or leaving a Trion on the surface, an excited state bound by two holes in the valence band \cite{Roncin_2002}.
As P. Rousseau was exploring similar processes with protons impinging the surface of a NaCl crystal, an unexplained structure showed up in the scattering profile and I suspected that it could be due to the part of the trajectory where the proton evolves as a $H^\circ$ atom. 
We decided to prepare a pulsed neutral beam of $H^\circ$ with our charge exchange cell and we recorded the first diffraction of keV atoms at a crystal surface\cite{Rousseau_2007}. 
Soon after the Berlin group of Pr. H. Winter also equipped with a position sensitive detector observed independently the diffraction of fast atoms \cite{Schuller_2007}, and we participated in the rebirth of our discipline.
So far, interactions of keV ions at surface were often considered as old physics since, apart from the atomic collisions itself, mainly classical mechanics was involved. 
By showing, as illustrated in fig.\ref{fig:diffract} that the crystal surface can act as a diffraction grating for the atomic wave several quantum features could be investigated, supernumerary rainbows \cite{Schueller_2008}, multiple scattering, bound state resonances\cite{Debiossac_PRL_2014}, van der Waals interactions...
Then, with M. Debiossac (PHD), after the identification of the elastic and inelastic diffraction, we managed to understand how a comparatively heavy projectile can bounce on the surface without transferring recoil energy to the surface atoms.
The answer was in textbooks, in a sudden approximation an harmonic oscillator can absorb a recoil  momentum $p$ without changing energy as in the Lamb-Dicke effect of recoilless emission of a photon. 
We could therefore show that both the projectile and the target surface are quantum objects and that bouncing on a surface has similarities with the Lamb-Dicke regime needed to trap cold atoms in an optical lattice \cite{Roncin_2017}.
In hindsight, it is obvious that the ability to build and calibrate time and position sensitive detectors \cite{Barat_2000,Villette_2000, Lupone_2015, Lupone_2018} adapted to specific needs was crucial. 

I would like to conclude by drawing attention to a recent result where Y. Picard, a former PHD student of Michel played a key role. 
In the group of D. Comparat at laboratoire Aim\'{e}-Cotton, he was involved in the construction of a custom combination of time and position sensitive detector with a clever strategy ; create a population of cold Rydberg $Xe^*$ atoms and photo-ionize them at threshold with a laser.
Then taking advantage of the impact location of the detected photo-electron to guess the position in space of the parent $Xe^+$ ion.
They can then, apply active corrections to the deflector along the path of the extracted ion \cite{Lopez_2019}.
In principle, such an ion beam where each ion is taken from a known place and actively guided to its desired destination is not restricted to Liouville conservation of phase-space and could pave the way to a new generation of experiments.

\section{conclusion}
Michel never hesitated to change field, when a new subject attracted his interest, he would wonder what is needed to contribute and try to initiate collaborations.
The examples are numerous were what he found was quite different from what was expected, but this is also part of the fun of doing Science, because sometimes, a bigger surprise was just waiting to be discovered.
"Try to anticipate, but never think that you know everything!", in other words, it is not because you imagine numerous obstacles that you should not go and look at unexplored situations.
Science is done by humans and it was a pleasure to do it with Michel and all his enthusiasm. 
Not that ha was always equal, certainly not, he had memorable explosions, but that was his way of expressing what was worrying him.
An hour later, everything would return to a quiet state without any resent.

\section{Acknowledgments}
The authors is grateful to all former members of the LCAM for their support and to A.J. Mayne for careful corrections of this manuscript.

\section{References}
\bibliographystyle{ieeetr} 
\bibliography{references}{}

\end{document}